# C-ITS bundling for integrated traffic management

**Evangelos Mitsakis[1], Areti Kotsi[1], Vasileios Psonis[1]**

[1]Centre for Research & Technology - Hellas (CERTH) - Hellenic Institute of Transport (HIT)
Email: emit@certh.gr, akotsi@certh.gr, psonis@certh.gr

## Abstract

Cooperative Intelligent Transportation Systems (C-ITS) enable vehicles' communication with each other (Vehicle-to-Vehicle, V2V) and with roadside infrastructure (Vehicle-to-Infrastructure, V2I). In the context of traffic efficiency, C-ITS technologies could assist in road network status visualization and monitoring, through data exchange, improving this way traffic control organization and traffic management implementation. Bundling is the provision of several C-ITS services as one combined service. The purpose of bundling is to harvest the usability of C-ITS services by developing a strategy for the operation and exploitation of services in real-time and within varying geographical areas. Two different dimensions of bundling have been recognized covering: 1) end-users, and 2) operators-managers. The objective of the operators-managers dimension is the integration of C-ITS services in operational traffic management. This work spotlights the operators-managers bundling dimension, presenting a framework based on a step-by-step approach for integrating C-ITS services in traditional traffic management.

## Περίληψη

Τα Συνεργατικά Ευφυή Συστήματα Μεταφορών (Σ-ΕΣΜ) επιτρέπουν την επικοινωνία μεταξύ οχημάτων (όχημα-όχημα) και μεταξύ οχημάτων και υποδομής (όχημα-υποδομή). Σχετικά με την αποδοτικότητα της κυκλοφορίας, τα Σ-ΕΣΜ βοηθούν στην οπτικοποίηση και παρακολούθηση της κατάστασης του οδικού δικτύου μέσω της ανταλλαγής δεδομένων, συμβάλλοντας στη βελτίωση της διαχείρισης και του ελέγχου της κυκλοφορίας. Ο όρος "bundling" αναφέρεται στην παροχή πολλαπλών υπηρεσιών Σ-ΕΣΜ μέσω μιας ενιαίας υπηρεσίας. Στόχος του bundling είναι να αξιοποιήσει τις υπηρεσίες Σ-ΕΣΜ σε μια στρατηγική για τη λειτουργία και την αξιοποίησή τους σε πραγματικό χρόνο και σε ποικίλες γεωγραφικές περιοχές. Στο πλαίσιο του bundling έχουν ορισθεί δύο διαφορετικές διαστάσεις: 1) τελικοί χρήστες, και 2) διαχειριστές της κυκλοφορίας. Σκοπός της διάστασης των διαχειριστών της κυκλοφορίας είναι η ενσωμάτωση των υπηρεσιών Σ-ΕΣΜ στη διαχείριση της κυκλοφορίας. Η συγκεκριμένη εργασία εστιάζει σε αυτή τη διάσταση και παρουσιάζει μια προσέγγιση βήμα-βήμα (step-by-step) για την ενσωμάτωση των υπηρεσιών Σ-ΕΣΜ στην κλασική διαχείριση της κυκλοφορίας.

***Λέξεις - κλειδιά:*** *Συνεργατικά Ευφυή Συστήματα Μεταφορών, bundling, διαχείριση της κυκλοφορίας, υπηρεσίες ευφυούς κινητικότητας*

## 1. Introduction

Since a few decades private sector business-to-consumer vendors like service providers, data aggregators and distributors and networks operators are offering mobility services to assist travelers in making their trips as comfortable and efficient as possible. As their success is growing and more travelers are using such mobility services, the influence of these services on travel behavior and traffic patterns in the road network (i.e. trip, mode, route, lane and speed choices) increases. The main challenge for road authorities resulting from this notion is how to incorporate or connect private sector mobility services in/ to day-to-day dynamic traffic management. Usually, in the context of traffic management, this is done in the light of larger objectives like accessibility, the environment and livability. Like more traditional traffic management measures like ramp metering, services can be "on" and "off" at certain times, at certain locations and to certain user groups. Hence, an operational traffic manager is most interested in those services offering "control variables" of some kind, as opposed to services which are continuous by nature (i.e. always "on").

Generally, the core tasks of a road operator/ traffic manager could be considered as ensuring the safe use of road infrastructure and smooth flow of traffic. In this context, a road operator/ traffic manager is considered mainly responsible for matters such as imposition of rules and prohibitions (i.e. maximum speed, opening and closing lanes, overtaking bans) and provision of information/ warnings/ advice (i.e. emergencies, traffic jams, alternative routes). The most common practice in traffic management centers around the world is the semi-automated support of response plans. A response plan is a procedural description that could be described by a set of production rules and is often represented as a flow chart. Each response plan deals with a specific situation like an event, rush hour or an incident at a specific location. Some decisions or actions in the response plan are automated and others need manual intervention (Spreeuwenberg & Krikke, 2017).

Nowadays traffic management is in a state of transition. Many innovations in traffic management deal with innovative road side equipment that work on a local level like an intersection or a combination of ramp metering and traffic lights (Hoogendoorn et al, 2016). Furthermore, neural network technology is used to optimize the cycle times of traffic lights resulting in new signal plans for different kinds of situations (Saraf, 1996).

From an operations point of view, it is expected that road operators/ traffic managers will gradually become an integral part of the network, relying on the alliances which would be formed with other parties, e.g. service providers, to shape traffic management. In this context there is a significant possibility for the private sector to actively contribute to the network by taking over traffic management tasks such as inform and advise road users. Towards this transition the integration of smart mobility concepts is considered as a combination of innovations which could assist in organizing and managing traffic in a better and more cost-efficient manner. Subsequently smart mobility services could promote this concept and enable vehicles to become an active part of managing traffic (Rijkswaterstaat, 2018).

This innovative perspective is addressed through this work which presents an approach to establish holistic traffic management by integrating "new generation" smart mobility services, hereafter denoted as C-ITS services, in a way which could promote collaborative synergies between traffic management centers and third parties. To practically implement this approach the bundling concept is introduced and thoroughly elaborated. The reminder of this paper is organized as follows: Section 2 presents and describes the bundling concept, Section 3 explains the methodology and analyses the theoretical background, Section 4 presents a case study where the proposed approach is applied in a city road network, and Section 5 provides conclusions as well as next steps for further research and implementation.

## *2. Bundling concept*

### *2.1 General description*

Bundling is the provision of several C-ITS services as one combined service. Within the bundling concept two different dimensions were recognized: 1) the End-users bundling dimension, and 2) the Operators-Managers bundling dimension. This work focuses on the Operators-Managers bundling dimension, hence a short description of the End-users bundling dimension is provided.

### *2.2 Description of the End-users bundling dimension*

In the context of the End-users bundling dimension C-ITS services bundles are developed and provided in the form of open, modular and extendable wrap applications (Apps) which bring together a complete suite of C-ITS services under one common user environment, with rich user experience features. The bundles are able to operate either in an automated mode, by providing context-, location- and user-preferences based information and guidance to the end user, as well as in a user-selected mode, where the end user selects the specific service or services relevant and useful for him/ her.

This bundling dimension differentiates between five major end-user types:
- Drivers, representing individual users of the road network and of services, using a motorized private vehicle.
- Vulnerable Road Users (VRUs), which are indicated as non-motorized road users, such as pedestrians and cyclists, as well as motor-cyclists and persons with disabilities or reduced mobility and orientation (European Union, 2010).
- Public Transportation Fleets' Operators.
- Commercial Fleets' Operators.
- Emergency services vehicles.

The main objective of the End-users bundling dimension is to provide integrated C-ITS through a single App, which could ease the widespread introduction of C-ITS services by enabling accessibility to large numbers of end-users. Anticipated benefits are safety increase, mobility improvement, environmental impacts decrease, and comfort.

The following table shows an indicative bundling of C-ITS services for each end-user type identified in the End-users bundling dimension.

*Table 1: Indicative bundled services per End-user type*

| C-ITS Services | Drivers | VRUs | PT | CF | EV |
|---|---|---|---|---|---|
| Rest-Time Management (RTM) | - | - | - | x | - |
| Motorway Parking Availability (MPA) | - | - | - | x | - |
| Urban Parking Availability (UPA) | x | - | - | x | - |
| Road Works Warning (RWW) | x | - | x | x | x |
| Road Hazard Warning (RHW) | x | - | x | x | x |
| Emergency Vehicle Warning (EVW) | x | - | x | x | x |
| Signal Violation Warning (SVW) | x | - | x | x | x |
| Warning System for Pedestrian (WSP) | x | x | x | x | x |
| Green Priority (GP) | - | - | x | - | x |
| Green Light Optimal Speed Advisory (GLOSA) | x | - | x | x | - |
| Cooperative Traffic Light for Pedestrian (CTLP) | - | x | - | - | - |
| Flexible Infrastructure (FI) | x | - | x | - | - |
| In-Vehicle Signage (IVS) | x | - | x | x | x |
| Mode & Trip Time Advice (MTTA) | x | x | - | - | - |
| Probe Vehicle Data (PVD) | x | x | x | x | x |
| Emergency Brake Light (EBL) | x | - | | - | - |
| Cooperative (Adaptive) Cruise Control (Urban CACC) | x | - | - | - | - |
| Slow or Stationary Vehicle Warning (SSVW) | x | - | x | x | x |
| Motorcycle Approaching Indication (MAI) | x | x | x | x | x |
| Blind Spot Detection/ Warning (BSD) | x | x | x | x | x |

*2.3 Description of the Operators-Managers bundling dimension*

Concerning the perspective of traffic management, the purpose of the Operators-Managers bundling dimension is to harvest the usability of C-ITS services by developing a strategy for the operation and exploitation of such services in real-time and within varying geographical areas. This bundling dimension differentiates between three different stakeholders:
- Road operators/ Traffic managers: When a traffic problem occurs on the network road operators are the first to find out and are responsible for solving this problem (for at least managing traffic). Nowadays, road operators use all the traditional dynamic traffic management services, e.g. display alternative route on Variable Message Signs (VMSs) they have available. To use these traditional traffic management services road operators do not depend on other stakeholders, because they own the assets needed for these services. When it comes to new services, e.g. Green Priority, also service providers are involved since service providers are the link between road operators and individual car drivers. Road operators should be able to turn green priority on and off to manage traffic in a sufficient way.
- Service providers (including car original equipment manufacturers (OEM)): To provide any kind of in-car information service providers should have a role in this process. For green light priority for example, the status (on/ off) should be provided to service providers so they can adapt their service (Green Priority) for their customers (end-users).
- End-users: The Operator-Manager bundling dimension is focused mainly on vehicle drivers (according to traditional dynamic traffic management). When new services as mode and trip time advice will be used, all travelers will be included.

Expected benefits/ impacts of the Operators-Managers bundling dimension are comprised of:
- Network optimization.
- Availability of more traffic management services for managing traffic, resulting in more effective operational traffic management.
- A common (interoperable) way for operational traffic management.
- Regulation of services. In others words, when two services together have a negative impact on traffic flow, road operators/ traffic managers should be able to decide which service to be active. Therefore, all service providers provide the same service(s) to road users at a specific geographical area at a specific time.

## 3. Methodology

The methodology proposed in this work is based on the approach for designing dynamic traffic management (DTM) "control strategies" (Krikke & Spreeuwenberg, 2015) and has been extended in a way to integrate C-ITS services in operational traffic management. As defined by the original approach a control strategy typically focuses on preventing congestion on road segments and on securing policy goals by optimizing traffic flows. It contains a framework including a step-by-step approach for selecting and activating DTM services. At present, traditional DTM services like providing more green time for a specific direction at an intersection and displaying alternative routes on VMSs mostly have been included in this framework, but it seems possible to incorporate smart mobility services as well. In this work the approach of integrating C-ITS services is further explored and elaborated.

### *3.1 DTM building blocks*

The DTM "control strategy" contains three main building blocks which help identifying DTM services that are useful for operational traffic management. The first building block is a policy defined by a road authority. A policy constitutes an important prerequisite of a traffic management control strategy. This includes a description of the importance and function of roads, as well as quantitative thresholds for links and route parts.

The second building block is the available road network defined by a road authority. In this context, different nodes, segments, links and route parts should be identified. The identification process is performed based on the following definitions:

- Choice nodes are nodes where the road user can choose between travel alternatives and can be used to affect traffic flow.
- Control nodes are nodes in the traffic network where the capacity of one or more directions can be affected and can be used to influence traffic flow.
- Regular nodes are intersections where traffic cannot be affected and road users do not have a choice between travel alternatives.
- Control segments are road sections in the traffic network where the capacity can be affected and can be used to influence traffic flow (e.g. flexible infrastructure, like a rush hour lane).
- Links are roads between two control nodes or between a control node and a choice node, and are used to detect traffic problems.
- Route parts are roads between two choice nodes and are used to detect traffic problems.

Once this building block is set, a road network like the one depicted in the following figure can be created.

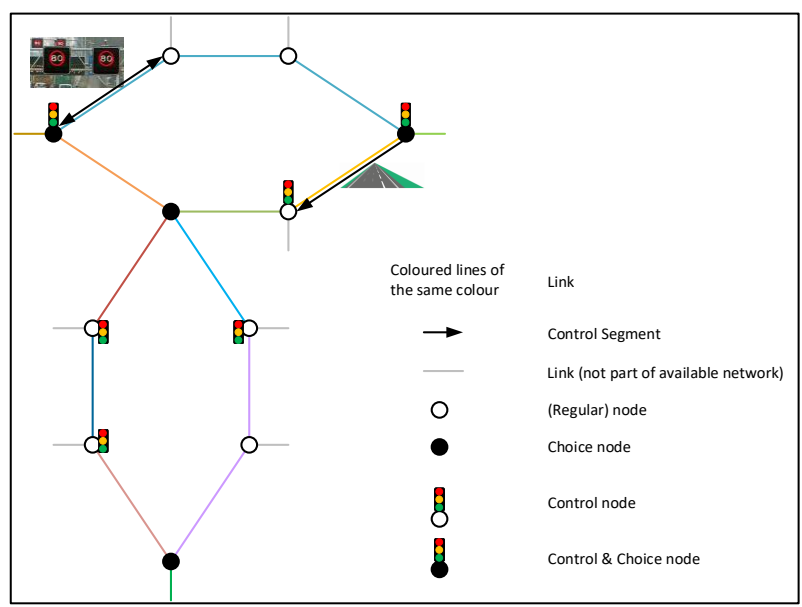

*Figure 1: Road network including links, choice nodes, control nodes and control segments*

The third building block is comprised of the strategies. Four different strategies are distinguished, each one containing one or more DTM services. Typically, but not necessarily, these strategies are applied as escalation phases, the first being the least severe one and the last being the most radical one:
- Inform traffic.
- Enlarge the outflow.
- Reduce the inflow.
- Reroute traffic (demand).

Once this last building block is set, it is necessary to predefine the available DTM services which are applicable to each choice node, control node and control segment, and can contribute to achieve each strategy. For traditional traffic management services, it is relatively easy to conceive how these can be activated or adapted within this framework. For example, simple changes to the configuration of a traffic light controller can enlarge the outflow or reduce the inflow of traffic at a certain control node. In this work, a process to introduce non-traditional smart mobility services, i.e. C-ITS services, offered by service providers is described.

### 3.2 Mechanism of integrated traffic management

Prior to implementing operational traffic management, a first step is to identify the contribution (in terms of impact) of smart mobility services to each of the four strategies (inform traffic, enlarge outflow, reduce inflow, reroute traffic). The following table presents the contribution of some indicative traditional traffic management services (TTM) and "new generation" smart mobility services (C-ITS).

*Table 2: Contribution of TTM and C-ITS services to strategies*

| Service | Service category | Primary objective | Inform traffic | Enlarge the outflow | Reduce the inflow | Reroute traffic |
|---|---|---|---|---|---|---|
| MPA | C-ITS | Publish information on facilities at and occupancy of truck parking areas | - | - | x | x |
| UPA | C-ITS | Publish information on facilities at and occupancy of truck parking areas | - | - | x | x |
| GP | C-ITS | Reduce delay time at traffic light for designated vehicles | - | x | - | - |
| FI | C-ITS | Control available road capacity | - | x | x | - |
| IVS (road section) | C-ITS | Present dynamic road sign information for road sections in | x | x | x | - |
| IVS (route) | | | the vehicle (personalized and extrapolated) | | | | |
| | TTM | Present route and travel time information in the vehicle (personalized and extrapolated) | x | - | - | x |
| MTTA | C-ITS | Multi-modal travel and departure time advice (MaaS-like concept, by incentives) | x | - | - | x |
| Shockwave damping | TTM | Inform drivers about appropriate velocity in traffic flow | x | x | x | - |
| Metering | TTM | Facilitate other traffic by metering | - | - | x | - |
| Modify traffic lights | TTM | Redistribute the green split at the intersection | - | x | x | - |

At this point it should be mentioned that not all C-ITS services are suitable for operational traffic management. For example, the service Emergency Brake Light has an impact on traffic safety, not on traffic management. Therefore, this service is not suitable for operational traffic management.

The second step is to identify the available C-ITS services and which apply to each choice node, control node and control segment. In our case this step differentiates from the logic of identifying smart traditional traffic management services where road authorities can identify the available traditional traffic management services themselves. For identifying C-ITS services road authorities are partially dependent on the C-ITS services provided by service providers. Therefore, road authorities should inform service operators which services are available.

Having completed the aforementioned actions, the operational process can be initialized. This process pertains certain similarities to the one used for activating traditional traffic management services. First of all, based on network assessment the traffic problem as seen from the traffic manager/ road authority perspective is identified and the traffic situation including bottlenecks and the primary causes of these bottlenecks are described. Thereafter, the preferred situation is defined, indicating the objective for the control strategy. The core step is to identify which (available) C-ITS services could provide (a part of) the solution for the traffic problem. For this purpose, it should be taken into account that some services are under direct control of a traffic manager, whereas others require the involvement of service providers and are therefore subject to business and collaboration models. The resulting set of (traditional and non-traditional) services together constitutes the control strategy to be activated to tackle the traffic problem.

An example of the operational process for activating C-ITS services is shown in the following figure. In this C-ITS service activation scenario a road operator detects a traffic jam in the network and needs to tackle this traffic problem. Therefore, the road operator chooses a strategy to enlarge the outflow by opening a peak-hour lane for drivers. The road operator provides this information to service providers. Service providers having an extra service available for drivers, i.e. Flexible Infrastructure, provide this to their customers. The customers of the service provider, i.e. end-users, receive this information on their personal devices (e.g. smartphones).

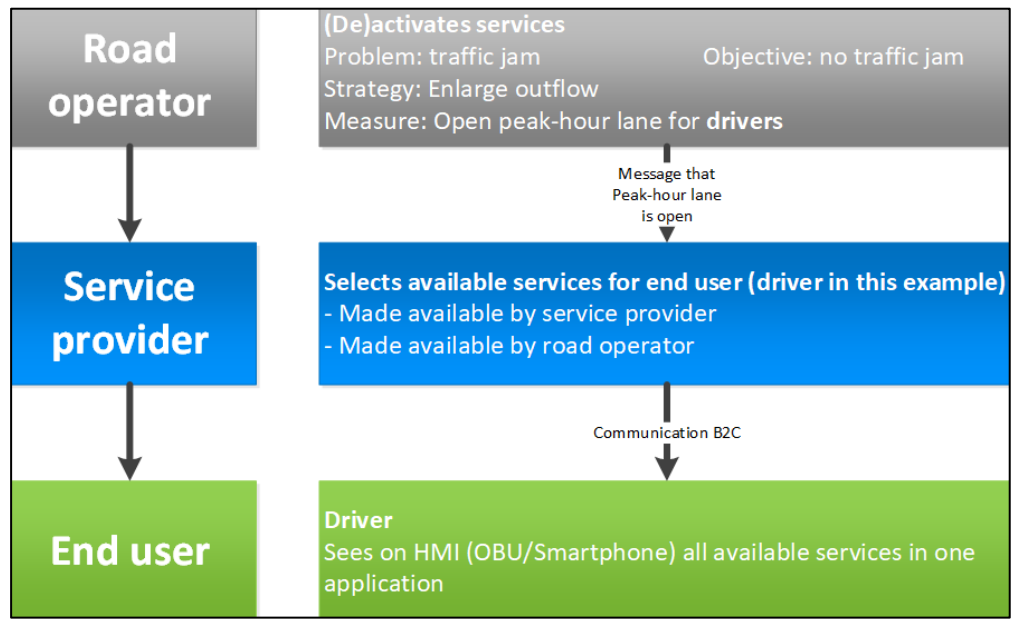

*Figure 2: Example of the operational process for activating C-ITS services*

To practically implement the mechanism of integrated traffic management, a step by step procedure has been defined, in order that road operators/ traffic managers can interpret the approach and fulfil all actions prior to the operational process. The step by step procedure is comprised of six questions which answers set the background for the operational process:
- Which C-ITS services are available in the area of interest?
- What is the deployment scale for each C-ITS service? (i.e. large-scale, limited scale, proof of concept)
- Who and how many are the end-users? (i.e. Drivers, VRUs, Fleets' Operators, Emergency services vehicles)
- Which is the available road network? (i.e. choice nodes, control nodes, choice & control nodes, regular nodes, control segments, links)
- Which are the common traffic problems in the available traffic network?
- How could each C-ITS service contribute to solving these traffic problems?

The application of the step by step process in thoroughly presented in the next section through a case study in the city of Thessaloniki, Greece.

## *4. Thessaloniki case study*

Thessaloniki is the second largest city in Greece covering a total of 1.455,68 km2 with an average density of 665,2 inhabitants per km$^2$. The total number of vehicles in the city exceeds 777.544, including private cars, heavy vehicles and motorcycles (Hellenic Statistical Authority, 2012). These characteristics in combination with the dense road network result in multiple mobility challenges which need to be addressed. Various actions have taken place in the past aiming to provide solutions through suites of services for travelers. Such services aimed to assist them in everyday mobility related decisions by providing real time mobility related and environmental conditions information, optimal route planning based on traveler-defined criteria (fastest, shortest, cost efficient and environmental friendly routing), public transport information and routing services, ride sharing and user awareness tools (Morfoulaki et al., 2011), (Mitsakis et al., 2015).

Recently an attempt to address the mobility challenges through the deployment of C-ITS services has been initiated as the city of Thessaloniki is a deployment site of the C-MobILE project which focuses on large-scale implementation and demonstration of C-ITS services in eight European cities (Barcelona, Bilbao, Bordeaux, Copenhagen, Newcastle, North Brabant, Vigo, Thessaloniki) (C-MobILE, 2019).

The most important prerequisite for the implementation of the mechanism of integrated traffic management was to define the city policy objectives. This called for the identification of services already existing in the road network, as well as future ones to be implemented. In our case, the focus was on the implementation of specific C-ITS services which would be in line with the city's transport and traffic policy objectives. Such objectives include solving/ mitigating the most common traffic problems, i.e. traffic congestion, parking problems and accidents. Hence, the city has set the goals to reduce accidents, increase safety, promote eco-friendly driving, provide real-time information, optimize traffic flow and increase traffic efficiency. C-ITS services, which could have a significant contribution, and their deployment scale were identified. Moreover, the actual recipients of these services were also identified by means of types and numbers of end-users.

*Table 3: C-ITS services available in Thessaloniki*

| Deployment Scale | C-ITS service |
|---|---|
|  | RWW |
|  | RHW |
|  | GLOSA |
|  | FI |
| Large-scale | IVS |
| Limited scale | WSP |
| Proof of concept | EVW |

|        |
|--------|
| SVW    |
| GP     |
| CTLP   |

*Table 4:* *End-user in Thessaloniki*

| End-user type | Number |
|---|---|
| Car drivers (individual users of the road network) | 6500 |
| Taxi drivers (commercial fleet) | 600 |
| Pedestrians (vulnerable road users) | Limited number |

As a next step the identification of the available road network, including a description of the importance and function of roads, its different nodes, segments, links and route parts, was performed. To proceed with this action, first an overview of all links and nodes of the city road network was created. Second, links and nodes which could be subject to adaptations in line with the strategies were identified, followed by a more thorough identification of major control nodes along specific links (see Figure 3). This network classification enables to precisely locate in which nodes and links the C-ITS services can be implemented, in order to activate targeted strategies and finally achieve policy objectives. The dynamic traffic management targets were based on the actual traffic problems (see Figure 4) and were aligned to the large-scale deployment C-ITS services which could have some form of impact (see Table 5).

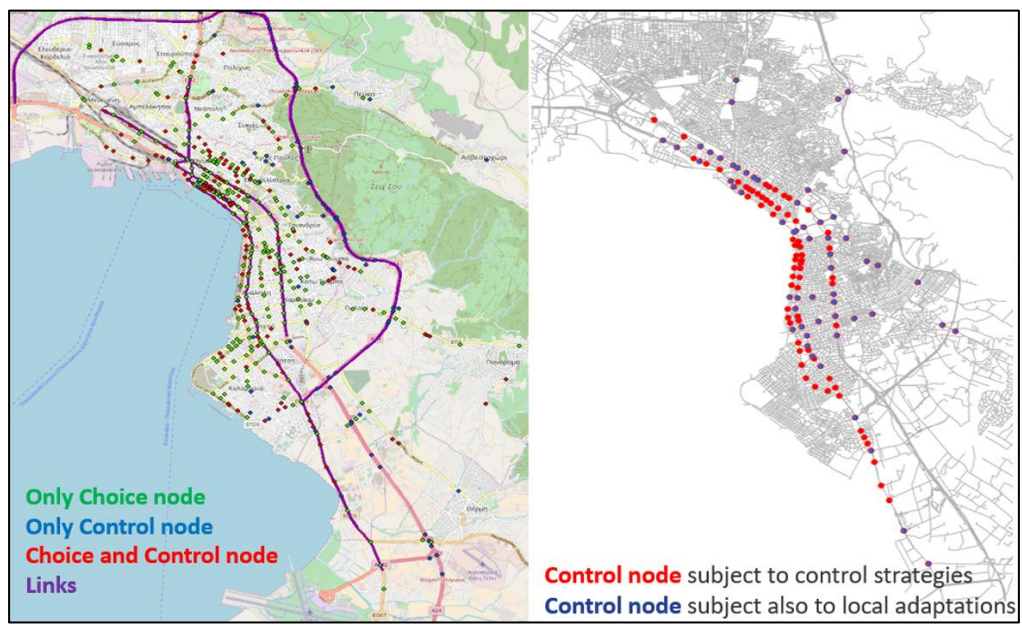

*Figure 3:* Available road network in Thessaloniki

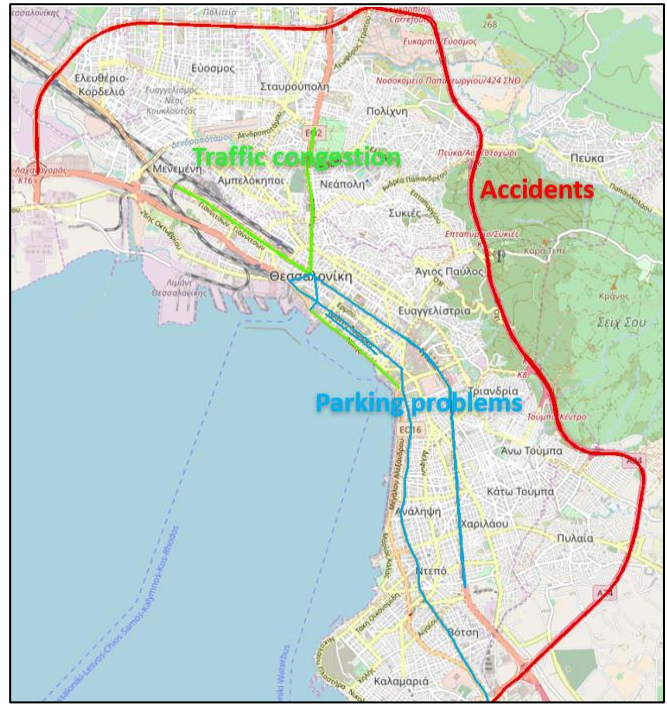

*Figure 4:* Traffic problems locations in Thessaloniki

*Table 5:* Contribution of Thessaloniki C-ITS services to strategies

| C-ITS Service | Inform Traffic | Enlarge Outflow | Reduce Inflow | Reroute Traffic |
|---|---|---|---|---|
| RWW | Inform about road works/ changes to the road layout. | - | - | - |
| RHW | Inform about hazardous locations downstream. | - | - | - |
| GLOSA | Inform vehicle drivers by providing speed advice/ green wave information. | - | - | - |
| FI | Inform about lanes status downstream. | More road capacity. | Less road capacity. | - |
| IVS | Inform about conditions, restrictions. | Speed harmonization. | Speed harmonization. | - |
| MTTA | Inform about conditions, restrictions and options. | Handles road capacity. | Handles road capacity | Delayed trips or trips by collective modes reduce demand. |
| PVD | (indirect effect) | (indirect effect) | (indirect effect) | (indirect effect) |

Since choice nodes, control nodes and control segments in the road network were identified, the next step was to predefine the available services which could contribute to achieving each strategy. The four aforementioned strategies represent different targets. As strategies are considered of being applied as escalation phases, the least severe one (Inform traffic) could be always applicable without influencing traffic and service providers, who could be allowed to constantly provide relevant C-ITS services. Moreover, it should be mentioned that in case of activating a specific strategy, relevant services can be activated along with services aligned to the less severe strategies. In the case of Thessaloniki, the C-ITS services were chosen to be implemented along all links where they could contribute to one of the strategies (see Table 6). For example, GLOSA was chosen to be implemented along links equipped with traffic lights at sequential nodes (intersections), where GLOSA could contribute to the "Inform traffic" strategy (see Figure 5: ).

*Table 6:* Mapping of C-ITS services in Thessaloniki to the available network

| Road Segment | Thessaloniki C-ITS services | | | | | | |
|---|---|---|---|---|---|---|---|
| | RWW | RHW | GLOSA | FI | IVS | MTTA | PVD |
| Choice Node | - | - | - | - | - | x | x |
| Control Node | - | - | x | - | - | x | x |
| Choice & Control Node | - | - | x | - | - | x | x |
| Regular Node | - | - | - | - | - | - | x |
| Control Segment | x | x | - | x | x | - | x |
| Link | x | x | - | x | x | - | x |

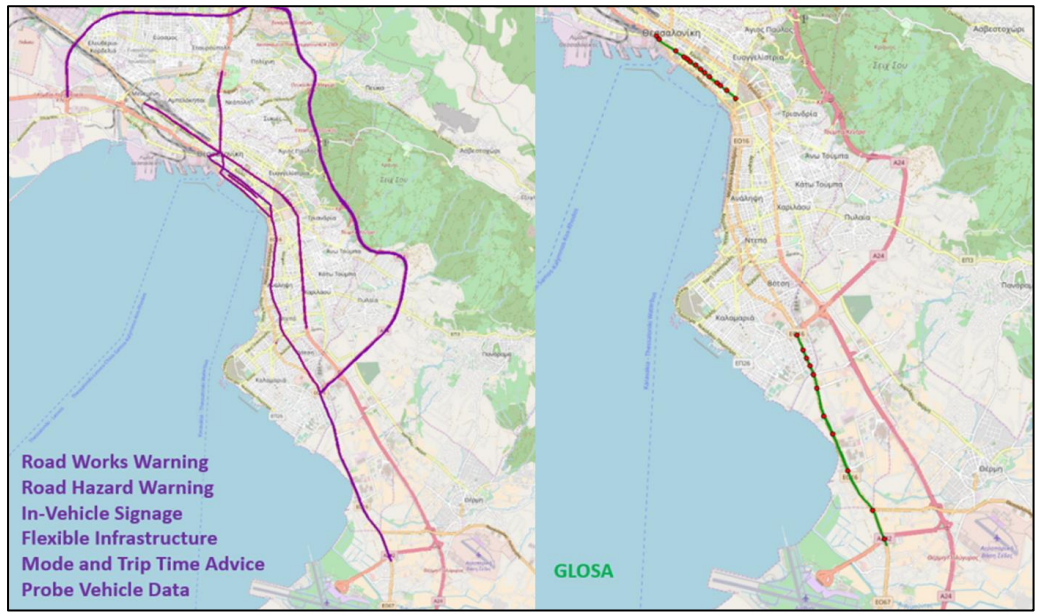

*Figure* **5**: C-ITS services applicable to the road network of Thessaloniki

Concerning the operational process, the necessary components and systems are under development for the city of Thessaloniki. These systems aim to enable the actual operation of a control strategy, where it must be possible for the traffic manager to monitor the entire available road network, the available C-ITS services (activated/ deactivated), and to choose the preferable strategy. In this framework, the collaboration between the traffic manager and a service provider is necessary so that the second is able to switch "on"/ "off" services. Afterwards, the information can be secure transmitted to the end-user, who will receive a message on his/ her personal information device.

## *5. Conclusions*

In this work an approach to integrate C-ITS services in operational dynamic traffic management by implementing control strategies is described. The integration of C-ITS services in day-to-day operational dynamic traffic management is expected to result in optimised network performance, as road operators/ traffic managers would be able to manage traffic better with more services being available. In addition, new partnerships could be established between road operators/ traffic managers and service providers promoting this way the collaboration among multiple stakeholders. Next steps to this work include the implementation of the operational process in the city of Thessaloniki, as actions targeting to the development and installation of technical components addressing the functional content of the approach are ongoing. Further research would be the development of impact assessment plans to evaluate the effects of the integration of smart mobility services, and more specifically C-ITS services, in day-to-day operational dynamic traffic management, both in terms of organizational aspects and traffic network performance.

## *6. References-Bibliography*